# Engineering high quality graphene superlattices via ion milled ultra-thin etching masks

**Authors:** David Barcons Ruiz[1], Hanan Herzig Sheinfux[1], Rebecca Hoffmann[1], Iacopo Torre[1], Hitesh Agarwal[1], Roshan Krishna Kumar[1], Lorenzo Vistoli[1], Takashi Taniguchi[2], Kenji Watanabe[2], Adrian Bachtold[1,3], Frank H.L. Koppens*[1,3]

[1] ICFO-Institut de Ciencies Fotoniques, 08860 Castelldefels (Barcelona), Spain
[2] Research Center for Functional Materials, National Institute for Materials Science, 1-1 Namiki, Tsukuba 305-0044, Japan
[3] ICREA-Institució Catalana de Recerca i Estudis Avançats, 08010 Barcelona, Spain

**Abstract**

Nanofabrication research pursues the miniaturization of patterned feature size. In the current state of the art, micron scale areas can be patterned with features down to ∼ 30 nm pitch using electron beam lithography. Our work demonstrates a new nanofabrication technique which allows patterning periodic structures with a pitch down to 16 nm. It is based on focused ion beam milling of suspended membranes, with minimal proximity effects typical to electron beam lithography. The membranes are then transferred and used as hard etching masks. We benchmark our technique by engineering a superlattice potential in single layer graphene using a thin graphite patterned gate electrode. Our electronic transport characterization shows high quality superlattice properties and a rich Hofstadter butterfly spectrum. Our technique opens the path towards the realization of very short period superlattices in 2D materials, comparable to those in natural moiré systems, but with the ability to control lattice symmetries and strength. This can pave the way for a versatile solid-state quantum simulator platform and the study of correlated electron phases.

## Introduction

Nanoscale fabrication is at the heart of the technological revolution of semiconductor technology and scientific breakthroughs in nanotechnology, progressing rapidly for more than five decades[1]. Accordingly, the push to further improve nanofabrication techniques is an ongoing effort to fit more electronic components per unit area or to develop quantum devices and quantum bits, relying on quantum coherent control in the nanoscale regime[2,3]. Another important application is in quantum materials, where materials' properties can be tuned in situ. Arrays of quantum dots with sufficiently large Coulomb interactions could lead to the observation of metal-Mott insulator quantum phase transitions and potentially even the d-wave superconducting phase[4]. Recent discoveries of correlated phases in twisted bilayer graphene due to the superlattice (SL) potential[5,6] motivate further exploration of graphene SLs with more versatility in the lattice design and in-situ control.

Nanofabrication techniques always involve a trade-off between patterning resolution and throughput. At one extreme is scanning tunneling microscopy, which can be utilized to rearrange materials on the atomic scale, but is also extremely slow. At the other is extreme ultraviolet optical lithography (EUV), the main lithography technique in the semiconductor industry. This type of nanofabrication can achieve small feature sizes[7,8] and is highly scalable, but is prohibitively complicated and expensive. EUV requires the use of a pre-fabricated hard mask, making the patterning expensive and inflexible.

In academic research, the most common nanopatterning technique is based on electron beam lithography (EBL), which can be thought of as a compromise offering reasonable throughput and nanoscale resolution. Forward scattered electrons limit the minimal patternable feature size, while secondary electrons due to (inelastic scattering) interactions with the resist limit the pitch resolution. As such, using high acceleration voltages and thin organic resists, patterns with 35 nm pitch can be obtained, but significant reduction of the feature size has not been demonstrated[9–11].

Another common nanofabrication approach uses focused ion beam (FIB) milling to pattern the target material directly. In particular, helium (He) FIB routinely achieves single digit single feature resolution and pitch on the order of 10nm[12,13]. However, two major limiting factors affect He-FIB-based nanofabrication: (1) Milling of materials directly on a substrate is accompanied by deposition of He bubbles inside the substrate, contaminating it and affecting surface topography. (2) Secondary collisions damage the quality and integrity of the patterned material. It is possible to reduce the damage by milling suspended materials (where substrate scattering is absent) or using resist lithography[14]. However, suspending the material is complex and the amount of FIB induced unintentional damage is still expected to modify the material properties in the form of ion implantation and/or amorphization[15,16]. Here, we propose and demonstrate a new nanofabrication technique that combines the extreme resolution of He FIB milling with minimal damage to the patterned material (Fig. 1).

Our technique uses ultrathin suspended hard masks that are placed in contact with the target substrate material. This alleviates proximity effects and feature broadening induced by scattering from the substrate. The suspended ultrathin mask can be subjected to harsh FIB patterning with minimal proximity effects and later transferred onto the target material to project the pattern by reactive ion etching (RIE). The resolution of our method is demonstrated directly by atomic force microscopy (AFM), whereas the quality is demonstrated by applying our technique to the application of SL potentials in graphene via patterned graphite electrodes, as shown in Fig. 2. Graphene could be very sensitive to the presence of charge traps, impurities and patterning errors in the patterned gate electrode. However, transport data from our samples indicates minor unintentional damage to the patterned electrodes, even for highly dense patterns with an 18 nm pitch.

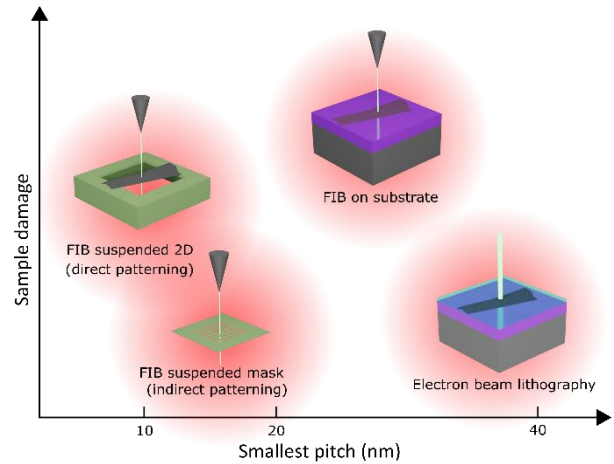

**Figure 1. Comparison chart of FIB alternatives to standard EBL for 2D materials patterning.** While FIB improves the resolution limit of EBL drastically, substrate damage has to be considered. Suspending the 2D material for the patterning process reduces substrate swelling dramatically compared to standard silicon wafer substrates but introduces disorder in the crystalline lattice. As an alternative, indirect FIB patterning of a suspended membrane, later used as an etching mask for the 2D material, results in no substrate damage and still preserves the high resolution of suspended FIB patterning.

Similar to moiré systems (e.g. twisted graphene[17] or graphene/hexagonal boron nitride (hBN) heterostructures[18,19]), it is possible to engineer SLs by applying a periodic electrostatic potential[20]. When applied to low-dimensional materials, such as graphene, this modifies its electronic band structure when the SL period is much smaller than the electronic coherence length[21]. Several approaches have been explored to introduce such potentials: from patterning the graphene directly[11], to inducing a periodic potential modulation by patterning the dielectric material [22,23] or the gate electrode [24–26]. In all cases, it is essential to reduce damage to the patterned material, keeping it flat and free of external contamination. However, to realize the full potential of artificial SLs, a new lithographic technique is required, one which can create high quality periodic patterning on the sub-20-nm scale, approaching the moiré length in graphene aligned on hexagonal boron nitride or magic angle twisted bilayer graphene.

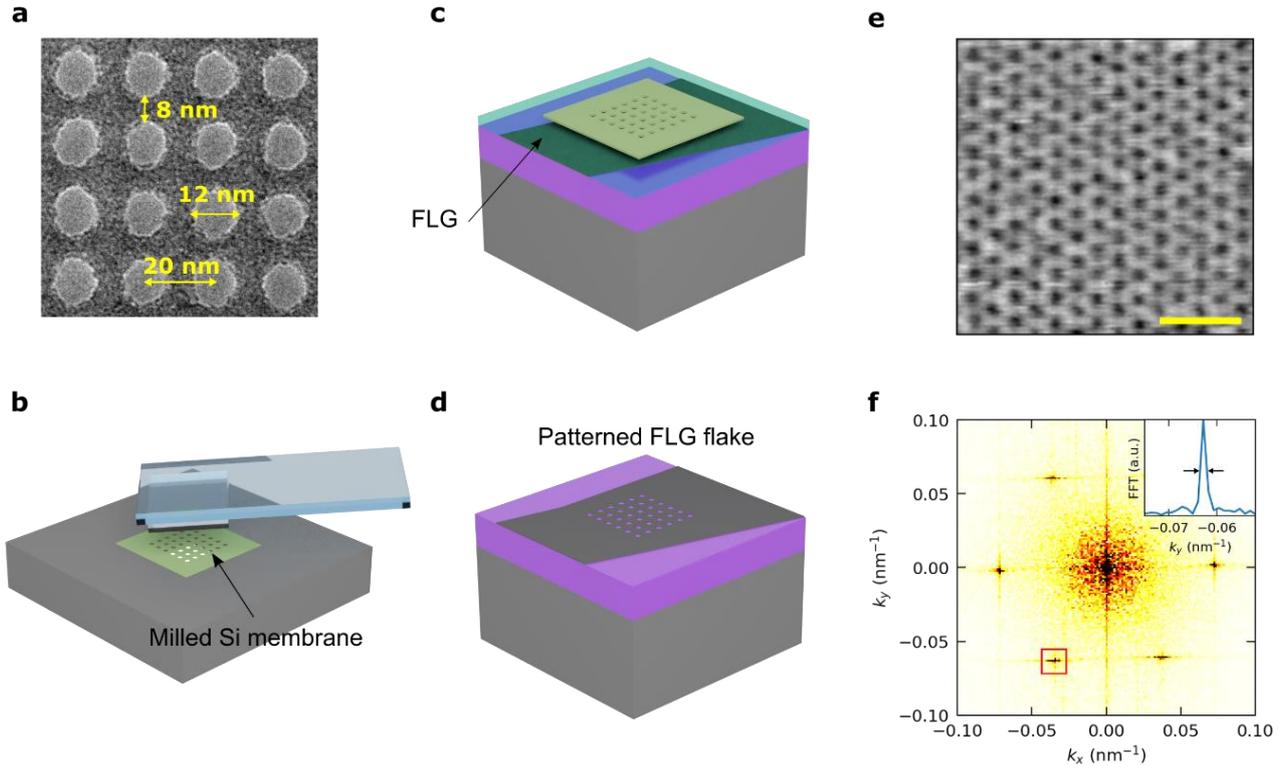

**Figure 2. Fabrication process of a patterned few layer graphene (FLG) gate electrode. (a)** Transmission electron microscope (TEM) image of a thin suspended silicon membrane which has previously been milled with a He-FIB **(b)** The membrane is transferred with a PPC/PDMS stamp onto a FLG flake coated with a thin layer of PMMA **(c).** The sample is etched following a standard $O_2$/Ar RIE process, followed by a lift-off process to remove the membrane and clean the PMMA layer underneath **(d). (e)** AFM topography image of a 16 nm period triangular lattice on a FLG flake. The scale bar is 50 nm. **(f)** FFT of a larger region of the same AFM image in panel (e), including 2555 lattice sites. Inset shows a vertical line cut on the area inside the red square.

### FIB indirect patterning of 2D materials

As the hard mask, we use a poly-crystalline silicon suspended ultrathin membrane, which is commercially available on a large scale and high quality. Importantly, the suspended membrane can be removed from its supporting frame and transferred with a polymeric stamp. The membrane is placed on top of a mechanically exfoliated target graphite flake, coated by a thin 22 nm layer of poly(methyl methacrylate) (PMMA). Due to the nanometric thickness of the hard mask, the aspect ratio is not a limitation and the feature size is only limited by the ion beam milling process. After the Ar/$O_2$ RIE process, we remove the hard mask with a standard acetone liftoff followed by vacuum annealing to remove polymer surface contamination. As a demonstration of the capabilities of our technique, we achieve a triangular lattice with a period as small as 16 nm and a hole diameter down to ∼ 8 nm (Fig. 2e).

To quantify the disorder of our lattices, we perform Fourier analysis of the AFM images (Fig. 2f), finding less than 3% period variations in our smallest lattices (16 nm period). By inspecting the transmission electron microscopy (TEM) images of our He-FIB milled silicon masks, we find a diameter of the holes of 13.2 ± 0.4 nm for a 22 nm period lattice, and thus the variation in diameter is only 2% and less than 6% in area. To showcase the high quality achieved with our technique, we incorporate a patterned graphite gate into two different SL graphene devices. Both are based on encapsulated single layer graphene with a square periodic lattice pattern in the bottom gate electrode (Fig. 3a), but with different period: 47 nm (Dev 1) and 18 nm (Dev 2).

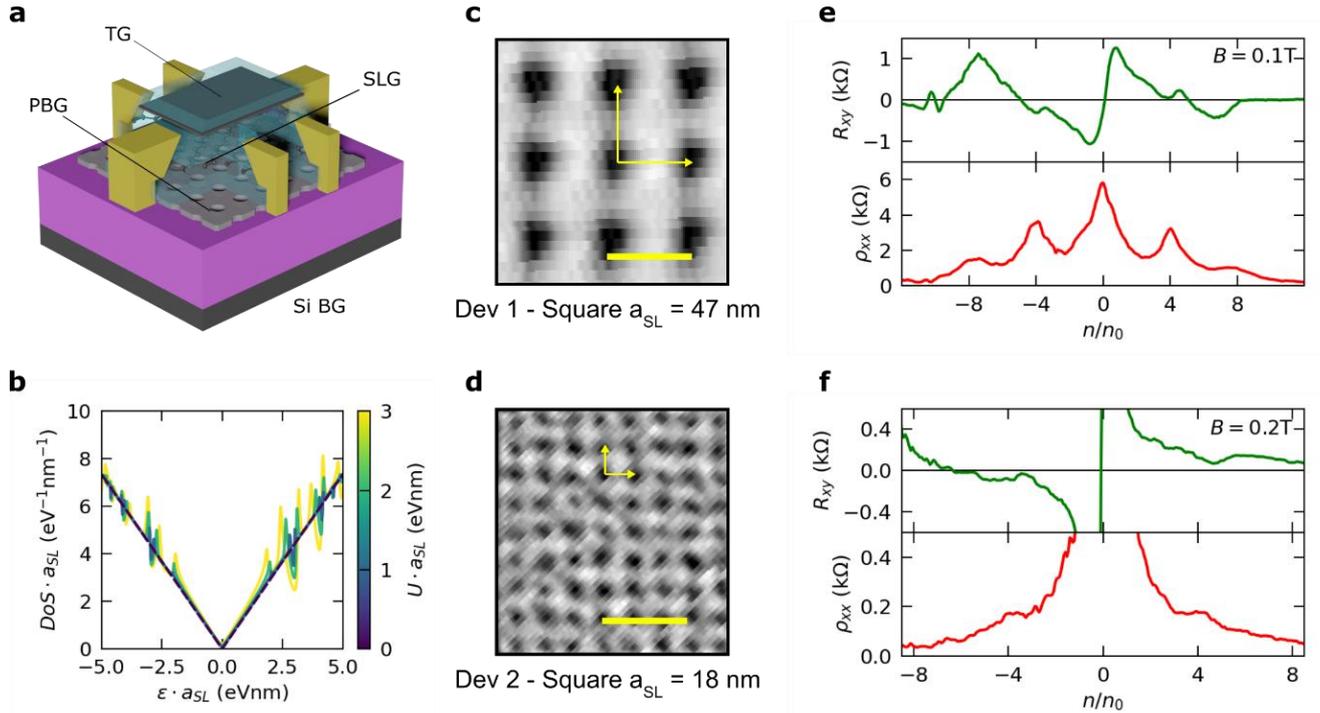

**Figure 3. Electronic transport characterization of devices Dev 1 and Dev 2. (a)** Device schematic. The main difference between Dev 1 and Dev 2 is the use of a graphite top gate in the latter. **(b)** Predicted normalized density of states per SL length, which demonstrates the invariance of the Dirac equation with the parameter $U \cdot a_{SL}$. **(c-d)** AFM topography images of the patterned graphite gates used for each device. The scale bars are 50 nm. **(e-f)** Longitudinal resistivity (red trace) and Hall resistance (green trace) as a function of the electron density per SL unit cell for Dev 1 ($V_{BG} = 70$ V) and Dev 2 ($V_{PBG} = 1.4$ V), respectively. Clear satellite peaks can be observed when the normalized density $n/n_0$ is a multiple of 4, with $n_0$ corresponding to the density where each SL unit cell is filled with one electron. This is consistent with the graphene´s four-fold degeneracy.

## Electronic transport characterization

Fig. 3c shows a zoom-in of the AFM characterization of the patterned gate electrode of Dev 1. The square SL in single layer graphene is expected to lead to the emergence of cloned Dirac cones equally spaced in energy due to band folding in the mini Brillouin zone[24]. By tuning the silicon backgate voltage (Si BG) and the patterned gate voltage (PBG), one can modulate the strength of the SL and the carrier type in the graphene layer to observe the cloned Dirac cones. In Fig. 3e, where the Si BG is kept at 70 V, two satellite peaks (indicating cloned Dirac cones) for electrons and another two for holes are clearly visible, being roughly of the main size as the main Dirac peak. The carrier density is normalized by the number of electrons per SL unit cell, $n/n_0$, and therefore, the spacing of the satellite indicates the four-fold (spin and valley) degeneracy in graphene. The clear observation of multiple satellite peaks, and their similar width and prominence compared to the main Dirac peak, demonstrate the high quality of the device and the high uniformity of the patterned gate. The second device studied in this work, Dev 2, has an 18 nm period square SL. To the best of our knowledge, there is no experimental work about electrostatically induced SLs with such a short periodicity. In this case, there is a top gate electrode (TG) as well, which can tune the carrier density of the graphene independently. On the other hand, the PBG only changes the carrier density on the areas of graphene right on top of the remaining graphite, while the areas on top of the patterned holes remain unaffected. This combination of gates allows us to control the SL strength and the total carrier density of the system independently. Therefore, we keep PBG fixed at 1.4 V and sweep TG to change the carrier density. In longitudinal resistivity and Hall resistance we observe a clear signature of the emergence of satellite peaks and an indication of carrier type sign change respectively, at $n/n_0 = \pm 4$ (Fig. 3f).

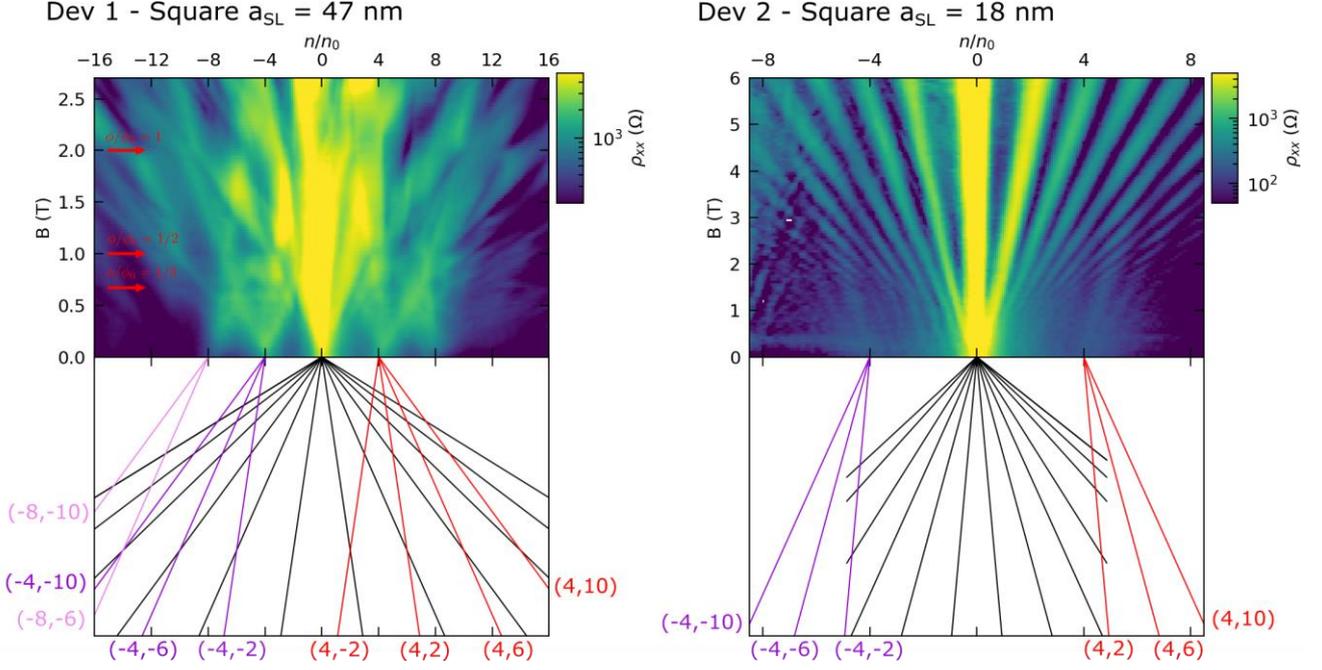

**Figure 4. Longitudinal magneto-resistance measurements of devices Dev 1 and Dev 2.** For Dev 1 ($V_{BG} = 70$ V), we observe a rich Hofstadter butterfly where Landau fans emerge from filling fractions $n/n_0 = 0, \pm 4, \pm 8$. Red arrows indicate resistance dips corresponding to an integer fraction of the flux quantum $\phi/\phi_0 = 1/3, 1/2, 1$. For Dev 2 ($V_{PBG} = 1.4$ V), Landau fans emerge from $n/n_0 = 0, \pm 4$. These are solid evidence of the formation of an 18 nm period SL. Colored lines below the magneto-resistance maps indicate the observed Landau fans. Labels indicate the SL filling fraction and Landau level ($n/n_0, LL$).

The prominence of the SL signature is different in the two devices. To understand this, we must examine the specific characteristics of both devices. As a rough approximation for the electrostatic potential modulation along the graphene layer for the formation of the SL, we consider a periodic one-dimensional system with a patterned bottom gate, global top gate (more details in supplementary information), and 50% duty cycle gate/vacuum. Due to the linearity of the Poisson equation, the induced modulation of the electrostatic potential $U$ scales with the ratio of the dielectric thickness $t_{BN}$ and the SL period $a_{SL}$. We estimate for Dev 1 ($t_{BN}/a_{SL} = 0.13$) $U \sim 200$ meV, while for Dev 2 ($t_{BN}/a_{SL} = 0.18$) $U \sim 120$ meV. Furthermore, we need to consider the scaling of the Dirac equation (Eq. 1). The larger the external potential term $U(x,y)$ is compared to the unperturbed graphene Hamiltonian $\mathcal{H}_0$, the stronger SL effects in the band structure will be:

$$\mathcal{H} = \mathcal{H}_0 + U(x,y)\mathbb{I}. \quad (1)$$

Since the momentum at the mini-Brillouin zone edge is inversely proportional to the SL period $a_{SL}$, decreasing $a_{SL}$ by a factor $\alpha$ enhances the unperturbed graphene term compared to the electrostatic potential term. In order to compensate for this enhancement, the external potential $U(x,y)$ would need to be larger by the factor $\alpha$. At the same time, the energy of the system will scale inversely proportional to $\alpha$. As a result, the comparatively small $U$ and small $a_{SL}$ in Dev 2 makes that the effect of the electrostatic modulation and, thus, the band structure modification in Dev 2 is weaker than that in Dev 1. Fig. 3b displays the importance of the Dirac invariant parameter $U \cdot a_{SL}$ (~9.4 for Dev 1 and ~2.2 for Dev 2) on the density of states.

Further evidence of the modified band structure is given in Fig. 4, where the gate voltage configurations are the same as in Fig. 3. For Dev 1 ($a_{SL} = 47$ nm), Landau fans emerge from the satellite peaks at filling fractions $n/n_0 = 0, \pm 4, \pm 8$. Given the relatively large period compared to typical moiré systems, it is possible to reach one quantum of magnetic flux inside the SL unit cell. We also observe Brown-Zak oscillations[27,28], indicated with red arrows corresponding to an integer fraction of the flux quanta $\phi/\phi_0 = 1/3, 1/2, 1$. For Dev 2 ($a_{SL} = 18$ nm), apart from the LL spectra emerging from charge neutrality, two other fans converge to densities $n/n_0 = \pm 4$. This is solid evidence for our device's 18 nm period SL formation.

## Conclusions

To conclude, we have successfully demonstrated a new nanofabrication technique that strongly alleviates proximity effects allowing to go beyond the spatial resolution limits of EBL. We used this technique to induce a SL potential in single layer graphene devices. To demonstrate the formation of the SL, we have presented electron transport data showing satellite peaks corresponding to cloned Dirac cones, Hofstadter butterfly spectrum, Brown-Zak oscillations and Landau levels emerging from the satellite peaks. To the best of our knowledge, the dimension of our smaller SL with $a_{SL} = 18$ nm sets a new record, enhancing by a factor of four the relevant Coulomb interaction strength $\propto 1/a_{SL}^2$ [11,22–24].

The ability to engineer arbitrary lattice geometries opens the path toward studying non-bipartite lattices[29] and flat bands in Dirac and gapped Dirac systems, such as the Lieb or Kagomé lattices, which require superior spatial resolution compared to patterned SLs achieved thus far. Furthermore, our technique enables a new generation of Fermi-Hubbard model simulators[30] when combining our patterned gate electrodes with 2D tunable semiconductors such as transitions metal dichalcogenides[31] or bilayer graphene[32]. Combining the patterned gate with a second gate allows the carrier filling and the Hubbard on-site interaction strength $U$ to be tuned independently. The superior quality of our nanopatterning process will make it possible to engineer lattices where the Hubbard $U$ can reach the 10-100 meV range, such that exotic correlated phenomena can be engineered at comparatively high temperatures.

## **Methods**

Detailed methods and extended discussion of techniques are available in the supplementary information.

Sample fabrication

The fabrication process of the patterned graphite gate is explained in the main text. After the cleaning process, we prepared an hBN/graphene/hBN (hBN/graphene/hBN/graphite/hBN in the case of Dev2) heterostructure. We first dropped it on a clean Si/SiO$_2$ substrate at a temperature of 158° C to clean its interfaces, and subsequently temperature was increased to 180° C to melt the poly (bisphenol A carbonate) (PC) film[33]. After dissolving the PC film in chloroform, the heterostructure quality is checked through AFM imaging and Raman spectroscopy. Finally, the heterostructure is again picked up and released following the same procedure on the patterned graphite gate. It is important to note that the hBN flakes between the graphene layer and the patterned gate electrode must be thin to allow for efficient doping modulation.

Standard EBL is used to pattern the Hall bar geometry and make one-dimensional edge contacts[34] to our heterostructure. In the case of the contacts, since they are right on top of the patterned graphite flake, we perform first an SF$_6$ etching process[35] to remove only the top hBN, followed by an O$_2$/Ar etching to remove the graphene, but not the thin bottom hBN that will prevent leakage current between the bottom graphite gate and the electrodes. We deposit Cr(3 nm)/Au(40 nm) followed by a lift-off in acetone.

Electronic transport measurements

Electrical measurements were performed in a He flow cryostat from ICE Oxford operating at T = 1.45K. Measurements were taken using standard lock-in techniques. A constant current of 5-20 nA was sourced by using a 10 MOhm resistor in series with our device at frequencies 11-18 Hz. Si BG, PBG, TG were controlled independently with a source meter.

AFM images processing

To remove substrate height variations, we apply a high pass filter with a cut off frequency $f = 1/3 \cdot f_{a_{SL}}$ by performing a 2D fast Fourier transform. See supplementary information for an example of the pre-processed and post-processed images.

**Data availability**

The data supporting this study is available from the corresponding author upon request.

**Acknowledgements**

D.B.R., H.H.S., R.H., I.T., H.A., R.K.K., L.V., A.B. and F.H.L.K. acknowledge funding from the Government of Spain through CEX2019-000910-S [MCIN/AEI/10.13039/501100011033], Fundació Cellex, Fundació Mir-Puig, and Generalitat de Catalunya through CERCA. D.B.R. acknowledges funding from the Secretaria d'Universitats i Recerca del Departament d'Empresa i Coneixement de la Generalitat de Catalunya, as well as the European Social Fund (L'FSE inverteix en el teu futur) - FEDER. H.H.S. acknowledges funding from the European Union's Horizon 2020 programme under the Marie Skłodowska-Curie grant agreement Ref. 843830. H.A. acknowledges funding from the European Union's Horizon 2020 research and innovation program under the Marie Skłodowska-Curie grant agreement no. 665884. R.K.K. acknowledges the EU Horizon 2020 program under the Marie Skłodowska-Curie grants 754510 and 893030. L.V. acknowledges funding from the H2020-MSCA-IF-2019 [887367-NanoMagnO]. A.B. acknowledges support from ERC Advanced Grant No. 692876 and MICINN Grant No. RTI2018-097953-B-I00, the European Union's Horizon 2020 research and innovation programme under the Marie Skłodowska-Curie grant agreement No. 847517 and 101023289, AGAUR (Grant No. 2017SGR1664), the Quantera grant (PCI2022-132951), the Fondo Europeo de Desarrollo, Recovery, Transformation and Resilience Plan-Funded by the European Union – NextGenerationEU and Quantum CCAA. F.H.L.K. acknowledges support by the ERC TOPONANOP (726001), the Government of Spain (PID2019-106875GB-I00), and Generalitat de Catalunya (AGAUR, SGR 1656). Furthermore, the research leading to these results has received funding from the European Union's Horizon 2020 under grant agreement no. 881603 (Graphene flagship Core 3) and 820378 (Quantum flagship).

**Author contributions**

All authors contributed to writing the manuscript. H.H.S. fabricated the silicon masks, and D.B.R. and R.H. fabricated the devices, with help from H.A. Measurements were taken by D.B.R. with help from R.K.K., L.V. and H.H.S. D.B.R. and H.H.S. performed the data analysis. T.T. and K.W. provided the hBN crystals. I.T. performed the electrostatics and band structure calculations. F.H.L.K. and A.B. supervised the project.

# Supplementary materials: Engineering high quality graphene superlattices via ion milled ultra-thin etching masks

**Authors:** David Barcons Ruiz[1], Hanan Herzig Sheinfux[1], Rebecca Hoffmann[1], Iacopo Torre[1], Hitesh Agarwal[1], Roshan Krishna Kumar[1], Lorenzo Vistoli[1], Takashi Taniguchi[2], Kenji Watanabe[2], Adrian Bachtold[1,3], Frank H.L. Koppens*[1,3]

[1] ICFO-Institut de Ciencies Fotoniques, 08860 Castelldefels (Barcelona), Spain
[2] Research Center for Functional Materials, National Institute for Materials Science, 1-1 Namiki, Tsukuba 305-0044, Japan
[3] ICREA-Institució Catalana de Recerca i Estudis Avançats, 08010 Barcelona, Spain

## Contents

## 1. Details of the membrane milling

Fabrication of the devices begins with commercially available polycrystalline silicon membranes (US100-A05Q00, SiMPore). Prior to milling in FIB, the membranes are annealed in a $H_2$:Ar atmosphere, at 400 C, for 3 hours. We find that this step of annealing greatly reduces mechanical strain in the Si membranes. On inspection in optical microscope, the membranes appear slightly wrinkled before the annealing process, whereas after annealing the membranes noticeably flatten.

The membranes are mounted on a hole in the sample holder, to improve imaging contrast and loaded into the He focused ion beam microscope (HIM). The HIM is then aligned to high precision on metallic features in the sample, with typical parameters being 5 µT He pressure, 10 or 20 µm aperture and spot control between 3 and 5. This yields a beam current between $2-12$ pA and resolution between $\sim 2-10$ nm (determined by visualizing sharp edges and features), with the exact milling parameters determined to match the achievable resolution with the desired periodicity of the pattern. Notably, the use of relatively high currents allows faster milling and reduces mechanical drift and similar undesirable effects.

After initial alignment, the membrane area is located in the HIM (the thin membrane shows clear contrast with the thicker, more conductive, background surrounding it). The sample is usually allowed to rest for 30-90 min (overnight when time allows), to minimize the effects of piezo drift in the sample stage. Following this resting period and additional finer alignment, the membrane is milled using a preprogrammed point-by-point deflection list. For large features, a milling dose comparable to $0.9$ nC$\mu$m$^{-2}$ is used. For smaller features, and especially when the lattice period approaches the HIM resolution (for the specific parameters used), the milling dose is optimized via demo patterns, where milling is typically sequential (single repeat rather than multiple repeats). The size of the milled area is typically kept on the order of $3 \times 3$ µm$^2$ in accordance with experimental needs. After milling the desired patterns, 20 nm wide cuts are made to the side of the membrane, leaving it partially connected to the frame via $3-4$ µm wide connecting bridges. These cuts facilitate breaking the membrane later during the transfer process.

In addition to the majority of the samples produced during our research, using He FIB, we were also able to produce similar results (but with larger milling periods) using a Gallium focused ion beam microscope. These samples were processed similarly to the above described samples milled by HIM, with the exception being that the membrane was mounted with the window facing down (situated above a hole in the sample holder), in order to facilitate locating the intended milling area with the FIB.

## 2. Details of the graphite gate patterning

In Fig. S1 we depict the transfer process of the silicon membrane and the etching of the graphite flake in detail. Moreover, we show in Fig. S2 optical and AFM images of every step in the fabrication process of a 30 nm period hexagonal lattice.

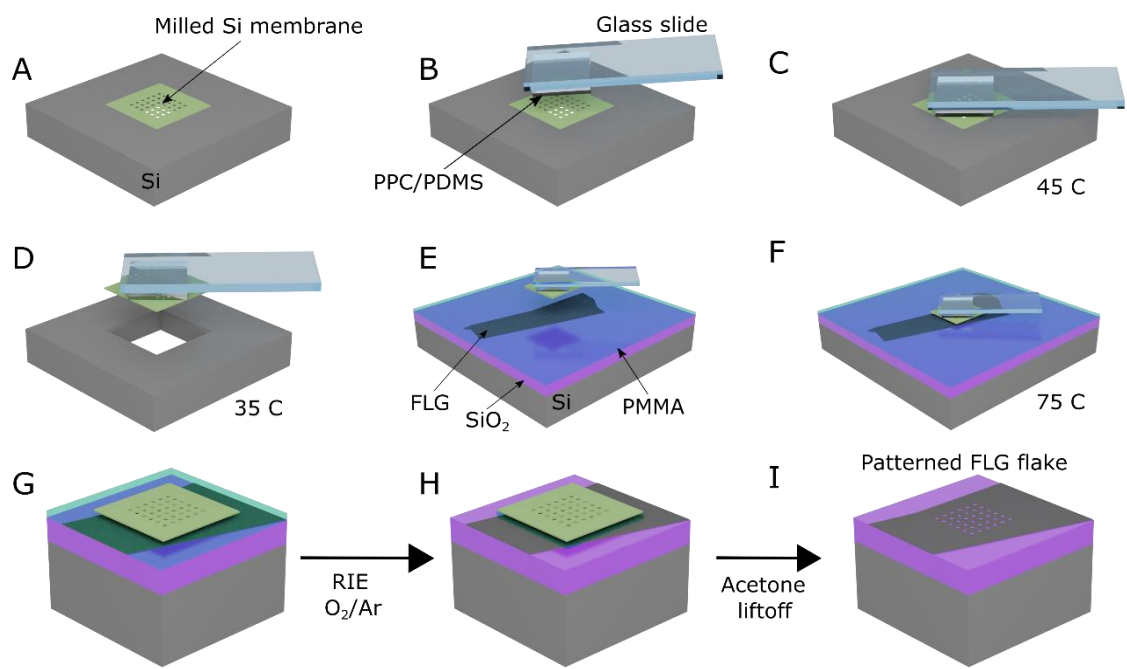

**Figure S1.** *The fabrication process of a patterned graphite gate.* **(A)** A suspended silicon membrane, as thin as 5 nm, is milled with FIB. **(B-D)** The membrane is detached and picked up from its supporting frame with a polymeric stamp made of PPC/PDMS[1]. **(E)** A Si/SiO$_2$ substrate with a few-layer graphene flake (FLG) is prepared in parallel, with a very thin layer of PMMA spin coated on the surface. **(F-G)** The milled membrane is released on the new substrate by heating up to decrease the adhesion to the PPC film. **(H)** A standard O$_2$/Ar RIE process is followed to transfer the pattern from the membrane to the FLG. **(I)** Finally, the sample is sonicated in acetone for a short time to clean the PMMA and remove the silicon membrane.

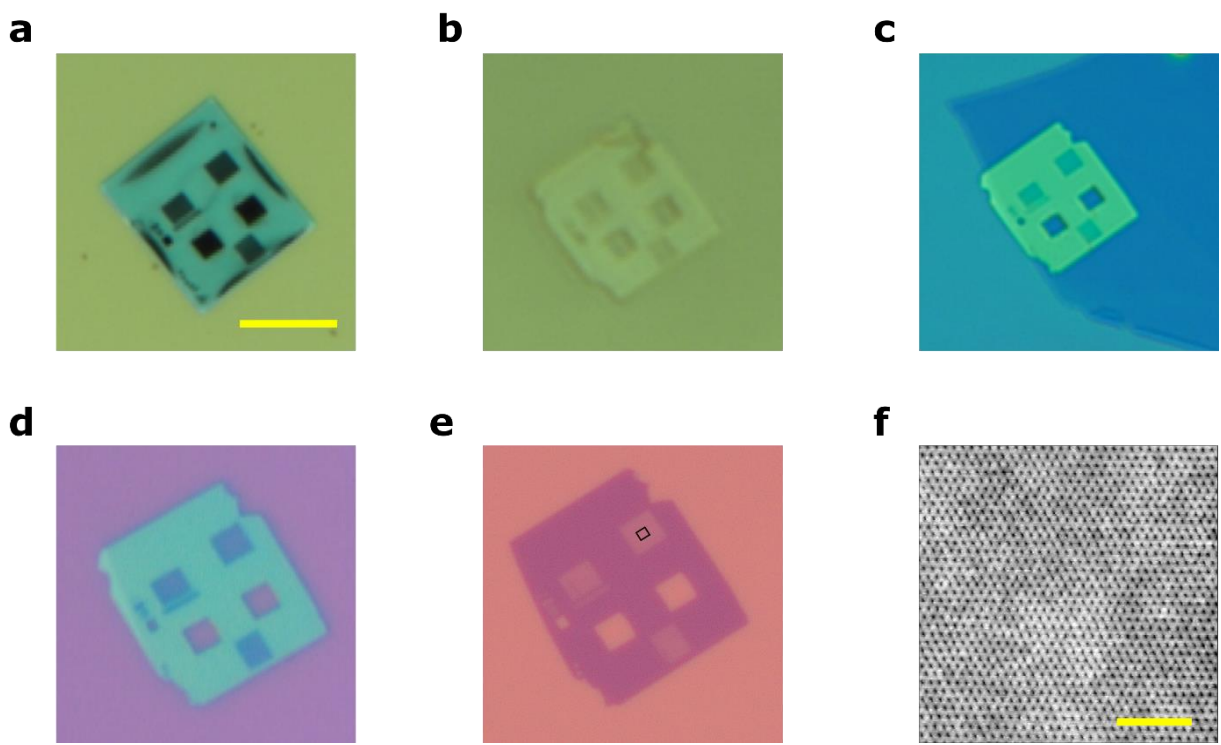

**Figure S2.** Fabrication of a 30 nm period hexagonal lattice on a FLG. (a) A 15x15 um$^2$ and 10 nm thin Si membrane is patterned with a He FIB. Scale bar is 10 µm (b) The membrane is picked up with a PPC/PDMS stamp and (c) dropped on the target FLG flake (around 3 nm thick). (d) An O$_2$/Ar RIE process is followed (20W, O$_2$/Ar, 40/40 sccm, 10 Pa, 1'40") and (e) finally the silicon membrane is removed by acetone sonication for 2 minutes, followed by 15 minutes of acetone cleaning plus 3 hours vacuum annealing at 600 C. (f) AFM topography image of the resulting 30 nm hexagonal lattice. Scale bar is 250 nm.

## 3. Limitations of the technique

There are some limitations that compromise the resolution limit when patterning highly dense lattices

- One of the main limitations is the mechanical stability of the suspended mask. Due to its nanometric thickness, milling a highly dense lattice where a big fraction of the material is removed results in holes merging or big areas breaking due to stress relaxation. In particular, producing patterns with very fine periods and large overall areas is further complicated by any minimal amount of drift or deformation in the long range order of the milled pattern, which cause neighboring holes (or hole rows) to overlap. These issues can largely be alleviated by using alternate custom-made membrane windows with thinner membrane thickness or by using mechanically stronger materials (e.g. non stochiometric SiN). Thinning the membrane contributes to reducing the milling dose, thereby reducing secondary ion collision damage and allowing the usage of higher resolution (lower current) HIM beam conditions.

- Related to the mechanical stability of the membrane is the aspect ratio of the patterned membrane – the ratio of the hole size (lattice period) to the membrane thickness. While He FIB milling is known to achieve high aspect ratios, we have observed that increasing the membrane thickness decreases the achievable resolution. This is most likely due to variations in the hole size between the top and bottom of the membrane (in addition to the increased aforementioned secondary ion collision damage).

- It is therefore expected that using dedicated Si membranes with thinner Si windows will enable further reduction in the minimal lattice period. Similarly, alternate membrane window materials can be used, which are mechanically stronger than polycrystalline Si (e.g. ultrathin non-stochiometric SiN). In this regard, it should be noted that while the membranes used are nominally 5 nm thick, the actual thickness measured in AFM (after transfer to substrate) is larger, on the order of 10 nm. Importantly, the membrane thickness (and the milling dose) does not change appreciably between different membranes (including those produced in different fabrication batches).

- Another part of the process which suffers from limited mechanical stability is the transfer process to the target substrate, where the membrane is picked up (broken) from the window frame and attached to a polymer. Due to the adherence properties of PPC, we change the temperature between the different steps in the transfer process. Even though the mask survives the transfer, we sometime observe long range disorder appearing in highly dense lattices due to stretching/compression of the mask. We attribute this to the thermal expansion of PPC. We expect that further optimization of the transfer process may reduce the number of such disordered lattices.

- As explained in the previous section, a buffer layer on the target substrate is needed to allow for a reliable lift off after etching. This results in a higher aspect ratio for the etching process, limiting the minimum feature size.

## 4. Electrostatic potential profile calculation

We estimated the electronic-density profile induced in graphene by the patterned gate by solving the equation of electrostatic using the Finite Element Method. For the sake of simplicity, we performed 2D simulations in a 1D-periodic geometry with the same lateral dimension of the real devices. We do not expect the solution of the full 3D equation to change the results qualitatively.

We considered the non-linearity due to the quantum capacitance of graphene, and we used the following values of the dielectric constants of $SiO_2$ ($\varepsilon = 3.9$) and hBN ($\varepsilon_{zz} = 3.56$, $\varepsilon_{xx} = 6.7$). We implemented the FEM using a Python code based on the open-source library FEniCS[2].

In Fig. S3 we show a side cut of Dev 1 and Dev 2, and their calculated electrostatic doping profile for a line cut crossing one hole. The voltage settings used are those that produce an average density $n/n_0 = -4$.

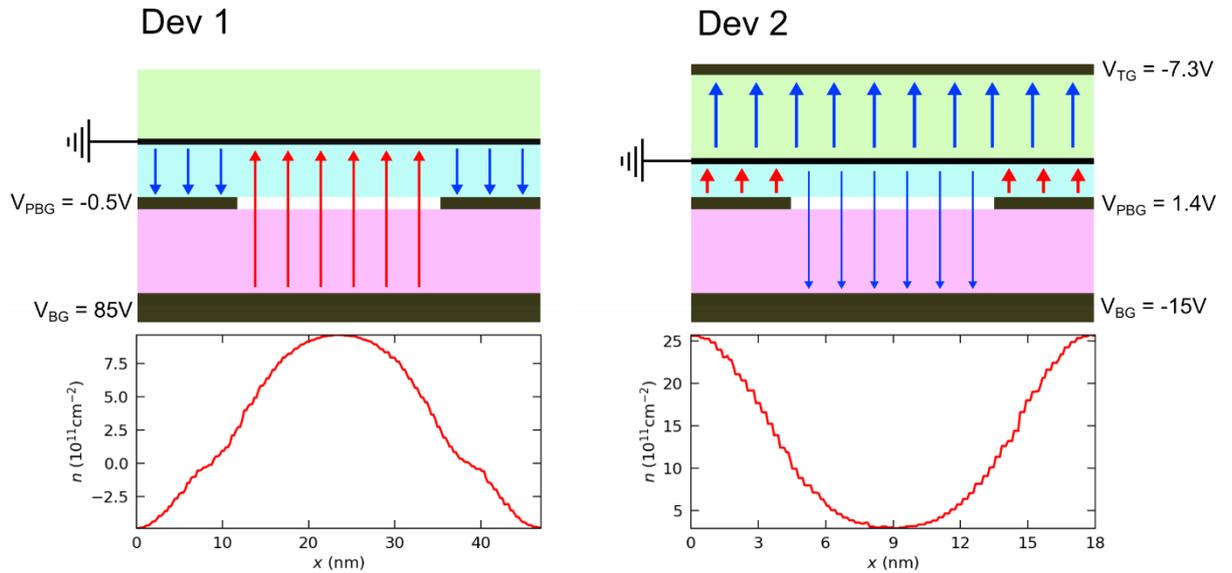

**Figure S3**. Induced carrier density modulation for $n/n_0 = -4$ voltage configuration for Dev 1 and Dev2. In both cases, the diameter of the holes is estimated to be $0.5 \cdot a_{SL}$ from the AFM topography images.

## 5. Band structure and conductivity calculation

We solve the following Hamiltonian for graphene under an external periodic scalar potential $U(\mathbf{r})$ in two dimensions

$$\mathcal{H} = v_D \boldsymbol{\sigma} \cdot \mathbf{p} + U(\mathbf{r})\mathbb{I}.$$

The scalar potential $U(\mathbf{r})$ is chosen to be a muffin-tin potential

$$U(\mathbf{r}) = u \sum_{\mathbf{t}} f_{MT}(|\mathbf{r} - \mathbf{t}|),$$

with $u$ being the potential amplitude and $f_{MT}(|\mathbf{r}|)$ representing a smooth potential well of radius $r_0$ and smoothing length $s$. The potential is expanded in a Fourier series with the reciprocal lattice vectors of the superlattice $\mathbf{G} = N_1 \mathbf{g_1} + N_2 \mathbf{g_2}$

$$U(\mathbf{r}) = \sum_{\mathbf{G}} U_{\mathbf{G}} e^{i\mathbf{G}\cdot\mathbf{r}},$$

with

$$U_{\mathbf{G}} = \begin{cases} u \dfrac{\tilde{f}_{MT}(G, r_0, s)}{\Omega_{u.c.}} & \text{if } \mathbf{G} \neq \mathbf{0} \\ 0 & \text{if } \mathbf{G} = \mathbf{0} \end{cases}.$$

Here, $\tilde{f}_{MT}(G, r_0, s) = \frac{2\pi r_0}{G} J_1(r_0 G) e^{-G^2 s^2}$ the Fourier transform of the well potential. We include enough $\mathbf{G}$ vectors such that $\hbar v_D G_{max} \gg \Delta U_{max}$

By adjusting the patterned hole radius $r_0$ and the smoothing parameter $s$, we are able to reproduce the potential profile calculated in the previous section.

In Fig. S4 we plot the calculated band structure for both devices discussed in the main paper, and the inverse Drude weight, which is proportional to the band conductivity.

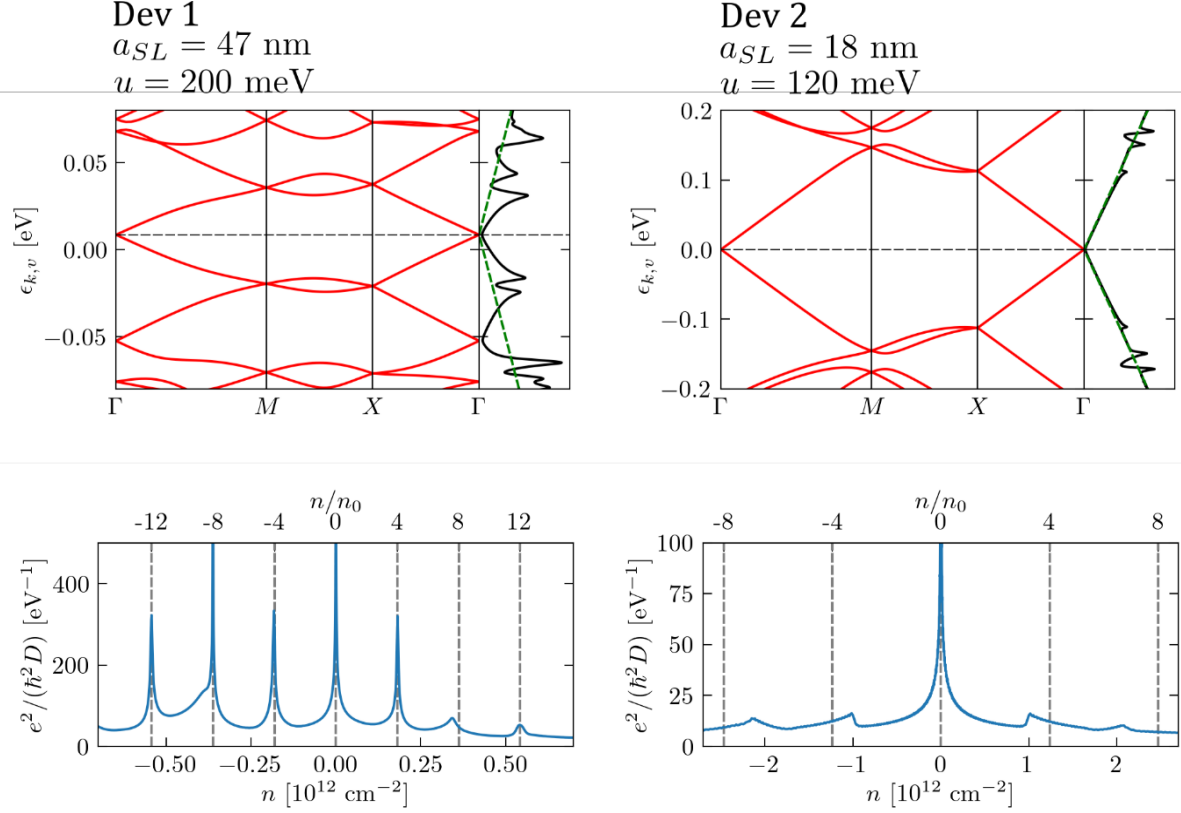

**Figure S4.** Calculated band structure, density of states and inverse of Drude weight for Dev 1 and 2. We consider the potential modulation for the calculation obtained from our electrostatics model. In the top panels, we plot the bands along the high symmetry point in the BZ of a square lattice, together with the density of states for the case with the modulation (black line) and the zero-modulation case (green dashed line). Bottom panels display the inverse of the Drude weight as a function of carrier density and filling fraction of the SL $n/n_0$

Importantly, the effective potential modulation induced in the graphene is also influenced by in-plane screening, to a degree that depends on the lattice period. Furthermore, the typical dependence of screening in monolayer graphene on the carrier density means that the experienced modulation will be smaller for the higher order Dirac peaks, which in general can have a bearing on interpreting experimental results.

To account for the in-plane screening, we consider the random phase approximation (RPA), similar to its application in [3]. The effective magnitude of the potential in the graphene layer, $u_{\text{eff}}$ is reduced by a factor of $1/\epsilon$, where

$$\epsilon = 1 - v_{G_0}\tilde{\chi}(G_0; \omega = 0),$$

with $G_0 = \pi/a_{SL}$, $\tilde{\chi}(G_0; \omega = 0)$ being the d.c. susceptibility of monolayer graphene and $v_q$ the Coulomb interaction potential

$$v_q = \frac{2\pi e^2}{q\bar{\epsilon}}\left[1 - \frac{e^{-2q\eta t_1} - 2e^{-2q\eta(t_1+t_2)} + e^{-2q\eta t_2}}{1 - e^{-2q\eta(t_1+t_2)}}\right],$$

with $t_1, t_2$ being the thicknesses of the top and bottom hBN spacers ($t_1 \to \infty$ for Dev 1 since there is no top gate) and $\bar{\epsilon} \approx 4.88, \eta \approx 1.37$ being the hBN permittivity parameters.

Fig. S5 shows the screening effect in our devices, where the potential reduction is slightly stronger for the shorter period device, Dev 2, than for Dev 1. The effect of the screening depends on the SL filling fraction, i.e. carrier density. However, the variation is relatively slow. For example, the induced potential does not change dramatically (in the RPA) between filling fractions 4 and 8.

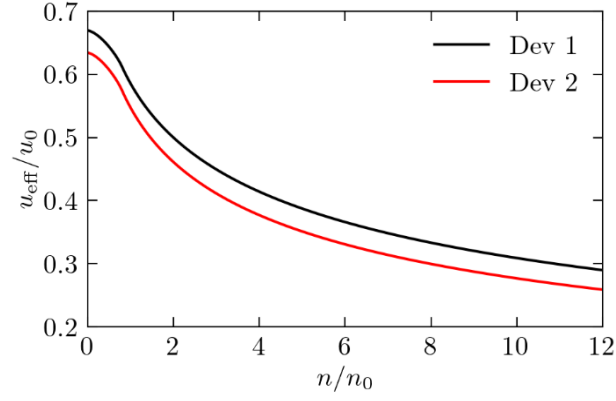

**Figure S5.** Effective electrostatic potential reduction $u_\text{eff}/u_0$ due to in-plane screening in the RPA, for Dev 1 and 2, as a function of the SL filling fraction $n/n_0$.

## 6. Longitudinal resistivity map for Dev 1

In Fig. S6 we show the emergence of the satellite peaks as a function of Si BG. The peaks appear as parallel lines to the main Dirac peak, and scale up in resistance as both gates are increased, i.e. towards higher modulation.

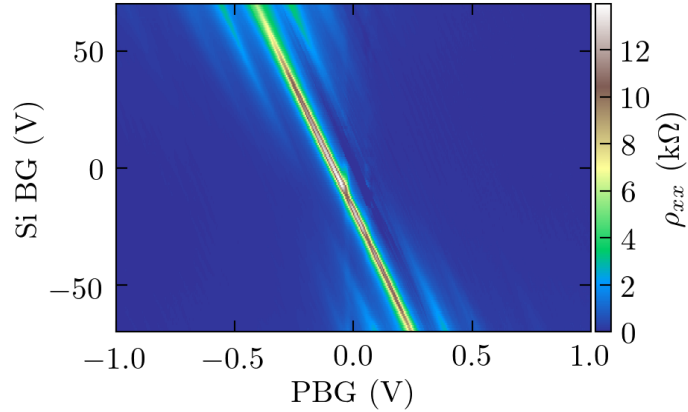

**Figure S6.** Dev 1 longitudinal resistivity as a function of patterned gate voltage (PBG) and silicon backgate (Si BG). Parallel resistance lines to the main one are the so called satellite peaks, which occur for multiples of $4n/n_0$.

## 7. AFM image filtering

Since AFM topography images suffer from substrate (SiO$_2$) height variations at the sub nanometer scale, we apply a high pass filter with cut off frequency $f = 1/3 \cdot f_{a_{SL}}$ by performing a 2D FFT. In Fig. S7 we show the AFM image from Fig. 1 before and after processing.

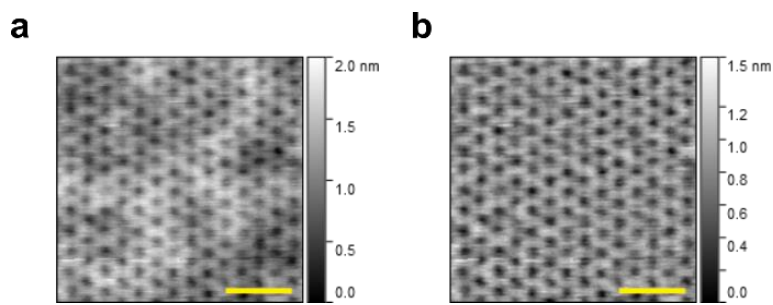

**Figure S7.** Unprocessed (a) and processed (b) AFM topography image of a triangular 16 nm lattice patterned on a graphite flake on a Si/ SiO$_2$ substrate. Scale bar is 50 nm.